\documentclass[aps,prb,twocolumn,preprintnumbers,amssymb,showpacs,superscriptaddress,floatfix]{revtex4-1}
\usepackage{amstext,amssymb}
\usepackage{graphicx}
\usepackage{xcolor}
\usepackage{amsmath}
\usepackage{amsfonts}
\usepackage{epstopdf}

\begin{document}
\title{Finite coupling effects in double quantum dots near equilibrium}

\author{Xiansong \surname{Xu}}

\affiliation{Department of Physics, National University of Singapore, Singapore
117551, Republic of Singapore}
\author{Juzar \surname{Thingna}}
\email{juzar.thingna@uni.lu}
\altaffiliation{Present address: Complex Systems and Statistical Mechanics, Physics and Materials Science Research Unit, University of Luxembourg, L-1511 Luxembourg, Luxembourg.}
\affiliation{Massachusetts Institute of Technology, Chemistry Department, Cambridge, Massachusetts 02139, USA}
\affiliation{Singapore-MIT Alliance for Research and Technology (SMART) Centre, Singapore 138602}

\author{Jian-Sheng \surname{Wang} }
\affiliation{Department of Physics, National University of Singapore, Singapore
117551, Republic of Singapore}

\date{\today}

\begin{abstract}
    A weak coupling quantum master equation provides reliable steady-state results only in the van Hove limit, i.e., when the system-lead coupling approaches zero. Recently, J. Thingna {\it et~al.} [Phys. Rev. E \textbf{88}, 052127 (2013)] proposed an alternative approach, based on an analytic continuation of the Redfield solution, to evaluate the steady-state reduced density matrix up to second order in the system-bath coupling. The approach provides accurate results for harmonic oscillator and spin-bosonic systems. We apply this approach to study steady-state fermionic systems and the calculation on an exactly solvable double quantum dot system shows that the method is rigorously valid up to second order in system-lead coupling only near equilibrium, i.e., linear response regime. We further compare to the Redfield and the secular Redfield (Lindblad-type) master equations that are inaccurate in all parameter regimes. Lastly, we consider the nontrivial problem of strong Coulomb interaction and illustrate the interplay between system-lead coupling, interdot tunneling, and Coulomb strength that can be captured only via the analytic continuation method.
\end{abstract}

\pacs{03.65.Yz, 05.70.Ln, 81.07.Ta}
\maketitle
\section{Introduction}
Quantum dots (QDs), also known as artificial atoms, are solid-state devices that confine electrons and exhibit a wide range of interesting phenomena. For example, the single-impurity or multiple-impurity Anderson models that map to QD systems have been extensively used to study Kondo physics\cite{Pustilnik2004,Cronenwett2004, KuboTokuraTarucha2008}. Various interference effects such as the Aharonov-Bohm effect\cite{Yacoby1994,Schuster1997,PhysRevLett.87.256802} and Fano effect\cite{Kobayashi2002,Lu2005} have also been observed in QD systems. They also form the building blocks of devices that exhibit negative differential resistance\cite{WunschBraunKoenigEtAl2005,PedersenLassenWackerEtAl2007,TrochaWeymannBarnas2009} and enhanced thermoelectric properties\cite{Wang2013b,Sierra2016}. 

The complete description of the QD system is encapsulated in the \emph{reduced} density matrix (RDM), i.e., the density matrix with the lead degrees of freedom traced out. The task of evaluating the RDM in a nonequilibrium setup is nontrivial due to the finite dissipative features of the system-lead coupling. The complexity further increases when the quantum dots are interacting via a nonlinear interaction like the Coulomb force. Typically, perturbative approaches on the Coulomb interaction based on Keldysh formulation \cite{Doyon2007} are used to tackle weak Coulomb interactions. Path integral \cite{Segal2011} based approaches can handle strong interactions, but are extremely complex and become numerically cumbersome with a large system Hilbert space. Other techniques such as the renormalization group methods\cite{Karrasch2010} evaluate only the important diagrams for strong Coulomb interactions and hence do not treat the Coulomb interaction exactly. Out of these numerous methods proposed, it turns out that the quantum master equation (QME) approach is the most suitable and efficient method to handle strong nonlinear interactions in QD systems.

A number of keystone QMEs have been formulated with the Nakajima-Zwanzig master equation\cite{Nakajima1958,Zwanzig1960} being the formally exact integrodifferential equation governing the evolution of the RDM. Despite its exactness, the equation is impossible to solve for general systems and hence a common simplification is to perturbatively expand the dissipative kernel. One example of such perturbative master equations is the Redfield master equation\cite{Redfield1957} (RME). The RME is a master equation formulated to incorporate the effects of weak system-lead coupling and could lead to unphysical negative populations\cite{Egorova2003,KulkarniTiwariSegal2013}. In order to avoid this drawback, one generally invokes a further secular approximation\cite{Breuer2007} that leads to a completely positive master equation of the Lindblad type\cite{Lindblad1976,Gorini1976}. One of the key obstacles with such perturbative QMEs is that they can accurately describe the steady-state RDM only when the system-lead coupling approaches zero\cite{Fleming2011a}. Hence in order to study finite coupling effects in the steady state, Thingna \emph{et~al.} proposed a method based on analytic continuation (AC) to capture effects at the second order in system-lead coupling. The approach was shown to be analytically valid for general system Hamiltonians in equilibrium \cite{Thingna2012} and numerically tested for harmonic and spin-bosonic systems in nonequilibrium \cite{ThingnaWangHaenggi2013,Thingna2014}. One of the key achievements of this approach is that it does not require the higher-order dissipative tensors\cite{Jang2002,Koller2010,Thingna2014} to obtain the enhanced accuracy and is computationally less cumbersome than the Redfield equation.

In this work, our first objective is to extend the analytic continuation technique to fermionic systems that possess a unique steady state \cite{Buca2012,Thingna2016} and test its validity. In order to achieve this objective, we numerically corroborate the AC technique with the nonequilibrium Green's function (NEGF) approach \cite{Dhar2012} for the exactly solvable spinless double quantum dot system. The AC approach provides an analytically exact solution, up to second order in the system-lead coupling in equilibrium, and in nonequilibrium the AC is numerically exact up to second order in the linear response regime. Despite its validity in the linear response regime, we show that the AC method surpasses the accuracy obtained by the Redfield master equation (RME) and the secular Redfield master equation (sRME). Our next objective is to study the effect of finite system-lead interaction for nonlinearly interacting systems near equilibrium. We achieve this by introducing a Coulomb interaction and show that the interplay between system-lead coupling, interdot tunneling, and Coulomb repulsion could lead to an enhancement or suppression of dot populations that cannot be captured by the standard weak-coupling quantum master equations.

This paper is organized as follows. In Sec.~\ref{Sec:II}, we present a double quantum dot (DQD) model and discuss the motivation for the choice of the observable. In Sec.~\ref{Sec:III}, the basic formalism of the analytic continuation technique for fermionic systems is outlined. In Sec.~\ref{Sec:IV}, the Coulomb interaction is introduced for spinless quantum dots and we study the interplay between the system-lead coupling, interdot tunneling, and the Coulomb repulsion strength.

\section{Nonequilibrium Double Quantum Dots}
\label{Sec:II}
\begin{figure}[t!]
\centering
\includegraphics[width=\columnwidth]{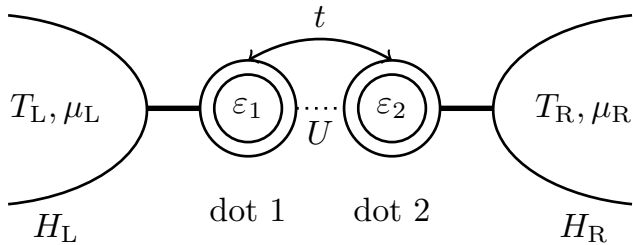}
\caption{A schematic illustration of a double quantum dot model. Dot $1$ is coupled to the left lead with temperature $T_{\mathrm L}$ and chemical potential $\mu_{\mathrm L}$ and dot $2$ is connected to the right lead. Electrons can hop between the two quantum dots with an interdot tunneling strength $t$ and there could be a presence of a nonlinear repulsive Coulomb interaction denoted by $U$.}
\label{fig:1}
\end{figure}

The total Hamiltonian for an open quantum system including the system of interest, leads, and system-lead coupling has the following generic form
\begin{eqnarray}
\label{eq:1}
H_{\mathrm{tot}} =H_\mathrm{S}+H_\mathrm{B}+H_\mathrm{SB},
\end{eqnarray}
where $H_\mathrm{S}$, $H_\mathrm{B}$, and $H_\mathrm{SB}$ describe the system, lead, and system-lead interaction Hamiltonians respectively. In this work we focus on the spinless double quantum dot system whose Hamiltonian reads
\begin{eqnarray}
\label{eq:2}
H_\mathrm{S}&=& \varepsilon_{1}\hat{n}_{1}+\varepsilon_{2}\hat{n}_{2}-t\left(d_{1}^{\dagger}d_{2}+d_{2}^{\dagger}d_{1}\right)+U\hat{n}_1\hat{n}_2,
\end{eqnarray}
with the two dots labeled by subscript $1$ and $2$ with energies $\varepsilon_1$ and $\varepsilon_2$ and number operators $\hat{n}_1=d_1^{\dagger}d_1$ and $\hat{n}_2=d_2^{\dagger}d_2$. The interdot tunneling strength and Coulomb interaction strengths are represented by $t$ and $U$, respectively, as illustrated in Fig.~\ref{fig:1}. We consider the leads to be infinite collections of free fermions with the Hamiltonian given by
\begin{eqnarray}
\label{eq:3}
H_\mathrm{B} = \sum_{k\in \mathrm{(L,R)}} \varepsilon_k c^{\dagger}_kc_k,
\end{eqnarray}
where $\varepsilon_{k}$ denotes the dispersion relation for the leads.

The system-lead coupling term is
\begin{eqnarray}
\label{eq:4}
H_\mathrm{SB}=\sum_{k\in\mathrm{L}}v_{k}d_{1}^{\dagger}c_{k}+ \sum_{k^\prime\in\mathrm{R}}v_{k^\prime}d_{2}^{\dagger}c_{k^\prime}+\text{H.c.},
\end{eqnarray}
with $v_{k}$ denoting the tunneling coefficients. In our nonequilibrium setup, the system and leads are coupled in such a way that dot $1$ is coupled with the left lead and dot $2$ is coupled with the right. The above system-lead coupling can be cast into a more general form given by
\begin{eqnarray}
\label{eq:5}
H_\mathrm{SB}=\sum_{\sigma} S_{\sigma}^1 \otimes B_{\sigma}^2+S_{\sigma}^2 \otimes B_{\sigma}^1,
\end{eqnarray}
where superscripts 1 and 2 denote the types of operators for both the system and the leads. $\sigma$ denotes the position of the lead, e.g., $\sigma=\mathrm{L}$ symbolizes the lead positioned at the left. In order to obtain the tensor product structure we perform a Jordan-Wigner transformation to the total Hamiltonian $H_{\mathrm{tot}}$ [Eq.~(\ref{eq:1})]. In the special case of the DQD model, the Jordan-Wigner transformation gives the transformed total Hamiltonian to be as the same form as Eq.~(\ref{eq:1}) except that the fermions are separately defined on system and leads. In other words, the transformed Hamiltonian is the same as the original with the fermionic nature ignored in the system-lead coupling \cite{Schaller}. Thus for the DQD model the system operators $S_{\sigma}$ take the form
\begin{eqnarray}
\label{eq:6}
S_\mathrm{L}^1&=&d_1~~~S_\mathrm{L}^2=d_1^\dagger,\nonumber \\
S_\mathrm{R}^1&=&d_2~~~S_\mathrm{R}^2=d_2^\dagger.
\end{eqnarray}
The corresponding lead operators are given by
\begin{eqnarray}
\label{eq:7}
B^1_\sigma =\sum_{k\in\sigma} v_{k}^* c_{k}~~~B^2_\sigma =\sum_{k\in\sigma} v_{k} c^\dagger_{k}. 
\end{eqnarray}

The tunneling coefficients $v_{k}$ can be characterized by the spectral density
\begin{eqnarray}
\label{eq:8}
\Gamma_{\sigma}\left(\varepsilon\right)=2\pi\sum_{k\in\sigma} |v_{k}|^2\delta\left(\varepsilon-\varepsilon_{k}\right)
\end{eqnarray}
that describes the properties of the lead. Throughout this work we will consider the spectral density to be a Lorentzian \cite{Wingreen1994},
\begin{eqnarray}
\label{eq:9}
\Gamma_{\sigma}\left(\varepsilon\right)=\frac{\lambda^2\Gamma_{\sigma}}{1+\left(\varepsilon/\varepsilon_\mathrm{D}\right)^2},
\end{eqnarray}
where $\varepsilon_\mathrm{D}$ is the cutoff energy and $\lambda^2\Gamma_{\sigma}$ is the overall effective system-lead coupling for the lead at position $\sigma$. In order to simplify our equations we will set $\hbar$ and $k_B$ as $1$. The dynamics of this model is studied via the numerically exact hierarchy equation of motion approach \cite{HartlePRB2014} and pronounced effects of the interdot tunneling strength are observed.

When the temperatures or the chemical potentials of the two leads are different, the DQD system $H_{\mathrm S}$ will be in nonequilibrium. In this situation, one of the most commonly observed quantities is the particle or heat current. The currents at the lowest order in the system-lead coupling depend only on the off-diagonal elements of the eigenbasis RDM (see Appendix A). In the case of perturbative master equations, since the steady-state off-diagonal elements are correct up to second order in system-lead coupling, the analytic continuation technique provides no added advantage. In other words, the currents evaluated via the analytic continuation and the Redfield equation would yield the same result for the lowest order of the currents. Thus, in order to explore the finite coupling effects related to both the diagonal and off-diagonal elements of the eigenbasis RDM, we constraint ourselves to the local populations of the dots,
\begin{eqnarray}
\label{eq:10}
N_{i} & =&\left\langle \hat{n}_{i}\right\rangle \nonumber\\
& =&\mathrm{Tr_S}\left[\hat{\rho}\hat{n}_{i}\right],
\end{eqnarray}
where $\hat{\rho}$ represents the reduced density operator.

The local population can be easily measured with various approaches such as quantum point contacts \cite{Fujisawa2006,Wang2013a,House2014} or quantum process tomography \cite{Yuen-Zhou2011}. Thus, we restrict our discussion from hereon to the dot population in order to study the interplay between system-lead coupling, interdot tunneling, and Coulomb interaction.
\section{Analytic continuation of master equation for fermionic systems}
\label{Sec:III}
The solution of a microscopic second-order perturbative quantum master equation is exact \cite{Fleming2011a} only in the limit when the system-lead coupling $\lambda^2\rightarrow0$. In other words, even though the solution contains all orders of $\lambda^2$ the correctness can be guaranteed only up to the zeroth-order coefficients. Thus, to obtain any higher-order effects, one needs to rely on the nontrivial higher-order quantum master equations \cite{Thingna2014}. In order to circumvent this obstacle, the analytic continuation technique was introduced that obtains the second-order populations from the second-order coherences \cite{Thingna2012}. We briefly outline the crux of the method in this section tailored to the DQD system.

We begin with the standard Redfield master equation expressed in the eigenbasis of the system Hamiltonian \cite{Redfield1957,Breuer2007,Blum2012}:
\begin{eqnarray}
\label{eq:11}
\frac{d\rho_{nm}}{dt} & = & -i\Delta_{nm}\rho_{nm}+\sum_{i,j}\left(\mathcal{R}_{nm}^{ij}+\mathcal{L}_{nm}^{ij}\right)\rho_{ij},
\end{eqnarray}
with $|n\rangle$ being the eigenstates of the system Hamiltonian, i.e., $H_{\mathrm S}|n\rangle = E_n |n\rangle$, and $\Delta_{nm}=E_n-E_m$ denotes the energy spacing between energy levels $E_n$ and $E_m$. Above the second-order relaxation four tensor only for the left-lead is given by
\begin{eqnarray}
\label{eq:12}
\mathcal{L}_{nm}^{ij} & =&  \sum_{\alpha,\beta = 1}^{2}\Big{[}S_{ni}^{\alpha}S_{jm}^{\beta}\left(W_{ni}^{\alpha\beta}+W_{mj}^{\alpha\beta*}\right) \\
&&-\delta_{m,j}\sum_{l}S_{nl}^{\alpha}S_{li}^{\beta}W_{li}^{\beta\alpha}-\delta_{i,n}\sum_{l}S_{jl}^{\alpha}S_{lm}^{\beta}W_{lj}^{\beta\alpha*}\Big{]}, \nonumber
\end{eqnarray}
with
\begin{eqnarray}
\label{eq:13}
W_{ij}^{\alpha\beta}&= & \int_{-\infty}^{t}  d\tau e^{-i\Delta_{ij}\left(t-\tau\right)}C^{\alpha\beta}\left(t-\tau\right).
\end{eqnarray}
Above $\alpha$ and $\beta$ denote the types of operators for both the system and the leads as defined in Eq.~(\ref{eq:6}) and Eq.~(\ref{eq:7}). $C^{\alpha\beta}\left(t\right)$=$\left\langle \tilde{B}^{\alpha}(t)B^\beta(0)\right\rangle$ is the lead correlation function with $\tilde{B}(t)=e^{iH_{\mathrm{B}}t} B e^{-i H_{\mathrm{B}}t}$ being the freely evolving lead operator. The lead correlators, when the leads consist of infinite number of free fermions, can be analytically evaluated for the Lorentz-Drude spectral density [Eq.~(\ref{eq:9})] as shown in Ref.~[\onlinecite{Zhou}].

Since the four tensor only pertains to the left lead, all the operators $S^{\alpha}$ and $W^{\alpha \beta}$ contain information about the dot $1$ (dot connected to the left lead) and the left lead, respectively. Therefore, the elements represented in Eq.~(\ref{eq:12}) would have a complete representation $S^{\alpha}_{ij}\equiv\left(S_{\mathrm L}^\alpha\right)_{ij}$ and $W^{\alpha\beta}_{ij}\equiv\left(W^{\alpha\beta}_{\mathrm L}\right)_{ij}$. Since we assume that the left and right leads are uncorrelated, the right lead four tensor $\mathcal{R}_{nm}^{ij}$ will have a similar form with $S^{\alpha}$ replaced by the dot $2$ operators [Eq.~(\ref{eq:6})] and $W^{\alpha\beta}$ replaced by the right lead information. To avoid the added notational complexity due to the two leads, the implicit summation over the leads will be presumed.

If there are no invariant subspaces, the steady-state condition $d\rho/dt=0$ will ensure that the system has a unique steady state. Thus, the equation governing the zeroth-order steady-state solution obtained in the van Hove limit ($\lambda^2\rightarrow 0$) reads
\begin{eqnarray}
\label{eq:14}
 \sum_{\alpha,\beta,i}\Biggl[W_{ni}^{\alpha\beta\prime} && S_{ni}^{\alpha}S_{in}^{\beta}-\delta_{n,i}\sum_{l}W_{li}^{\beta\alpha\prime}S_{nl}^{\alpha}S_{li}^{\beta}\Biggr]\rho_{ii}^{(0)}=0, \nonumber \\
&&  \rho_{nm}^{(0)}=0,~~~(n\neq m),
\end{eqnarray}
where $W_{nm}^{\alpha\beta}=W_{nm}^{\alpha\beta\prime}+iW_{nm}^{\alpha\beta\prime\prime}$. The above equation has the same form as the Pauli master equation \cite{Pauli} or the Davies form \cite{Davies} in the steady state. Following Ref.~[\onlinecite{ThingnaWangHaenggi2013}] an order-by-order method allows us to extract the second-order off-diagonal elements of the eigenbasis RDM as,
\begin{eqnarray}
\label{eq:15}
\rho_{nm}^{(2)}  &=& i \sum_{\alpha,\beta,i}\frac{S_{ni}^{\alpha}S_{im}^{\beta}}{\Delta_{nm}}\Big[\left(W_{im}^{\beta\alpha}\rho_{mm}^{\left(0\right)}+W_{in}^{\beta\alpha*}\rho_{nn}^{\left(0\right)}\right)\nonumber \\
& &-\left(W_{ni}^{\alpha\beta}+W_{mi}^{\alpha\beta*}\right)\rho_{ii}^{\left(0\right)}\Big],~~~(n\neq m).
\end{eqnarray}

The second-order diagonal elements of the eigenbasis RDM require the fourth-order relaxation tensor\cite{Thingna2012}, but in the analytic continuation method we try to obtain these elements via the second-order off-diagonal elements described in Eq.~(\ref{eq:15}). We achieve this by treating $\rho_{nm}^{(2)}$ as a function of $\Delta_{nm}$ and treat $\Delta_{nm}$ to be infinitesimally small such $\rho_{nm}^{(2)}\rightarrow\rho_{nn}^{(2)}$. This limiting value obtained via analytic continuation is given by
\begin{eqnarray}
\label{eq:16}
\rho_{nn}^{(2)} & = &  \sum_{\alpha,\beta,i} S_{ni}^{\alpha}S_{in}^{\beta}\Big[\left(V_{ni}^{\alpha\beta\prime\prime}\rho_{ii}^{(0)}-V_{in}^{\beta\alpha\prime\prime}\rho_{nn}^{(0)}\right) \nonumber\\ 
& &+W_{in}^{\beta\alpha\prime\prime}\frac{\partial\rho_{nn}^{(0)}}{\partial E_{n}}\Big].
\end{eqnarray}
The term $\partial\rho_{nn}^{\left(0\right)} / \partial E_{n}$ can be determined via Eq.~(\ref{eq:14}) by taking partial derivatives with respect to $E_{n}$ on both sides. The result takes the form 
\begin{eqnarray}
\label{eq:17}
\frac{\partial\rho_{nn}^{(0)}}{\partial E_{n}} =  \frac{\sum_{\alpha,\beta,i\ne n} S_{ni}^{\alpha}S_{in}^{\beta}\left(V_{ni}^{\alpha\beta\prime}\rho_{ii}^{(0)}+V_{in}^{\beta\alpha\prime}\rho_{nn}^{(0)}\right)}{\sum_{\alpha,\beta,i\ne n} W_{in}^{\beta\alpha\prime}S_{ni}^{\alpha}S_{in}^{\beta}}.
\end{eqnarray}
The elements $V_{nm}^{\alpha\beta}=\partial W_{nm}^{\alpha\beta}/\partial\Delta_{nm}$ and the prime and double-prime super-scripts describe the real and imaginary parts respectively. A foundational assumption is made such that $\rho_{nn}^{(0)}$ depends only on $E_n$. Normalizing the RDM one obtains the final form of the second-order diagonal elements of the eigenbasis RDM as
\begin{eqnarray}
\label{eq:18}
\rho_{nn}^{(2)} & = & \sum_{\alpha,\beta,i} S_{ni}^{\alpha}S_{in}^{\beta}\Big[\left(V_{ni}^{\alpha\beta\prime\prime}\rho_{ii}^{(0)}-V_{in}^{\alpha\beta\prime\prime}\rho_{nn}^{(0)}\right)+W_{in}^{\alpha\beta\prime\prime}\frac{\partial\rho_{nn}^{(0)}}{\partial E_{n}}\Big] \nonumber \\
&& -\rho_{nn}^{(0)} \sum_{\alpha,\beta,i} S_{ij}^{\alpha}S_{ji}^{\beta}W_{ji}^{\alpha\beta\prime\prime}\frac{\partial\rho_{ii}^{(0)}}{\partial E_{i}}.
\end{eqnarray}
Thus, Eqns.~(\ref{eq:14}),~(\ref{eq:15}), and~(\ref{eq:18}) are collectively termed as the \textit{modified Redfield solution} (MRS) and allow us to evaluate the RDM up to order $\lambda^2$, i.e., $\rho_{\mathrm{MRS}}=\rho^{(0)}+\lambda^{2}\rho^{(2)}$, without the contamination of higher-order inaccuracy that is difficult to characterize and predict.

In the same spirit that we obtained the second-order off-diagonal elements from the RME, we extract the incorrect diagonal second-order elements $\varrho^{(2)}$ by solving
\begin{eqnarray}
\label{eq:18.5}
\sum_{i}\left(\mathcal{R}_{nn}^{ii}+\mathcal{L}_{nn}^{ii}\right)\varrho_{ii}^{(2)}
& = & -\sum_{i,j \atop (i\ne j)}\left(\mathcal{R}_{nn}^{ij}+\mathcal{L}_{nn}^{ij}\right)\rho_{ij}^{(2)}.
\end{eqnarray}
Above $\rho_{ij}^{(2)}$ is the second-order off-diagonal elements obtained via Eq.~(\ref{eq:15}). The set of equations above to determine $\varrho^{(2)}$ are underdetermined and additionally require the normalization condition $\mathrm{Tr}[\varrho^{(2)}] = 0$.  Clearly, a direct analytic comparison between the MRS $\rho^{(2)}$ and RME $\varrho^{(2)}$ becomes impossible due to the complex structure of Eq.~(\ref{eq:18.5}).

In equilibrium, the RDM for fermionic systems takes the generalize grand canonical form
\begin{equation}
\label{eq:19}
\rho^{\mathrm{eq}}=\frac{\mathrm{Tr_B}\left[e^{-\beta \left(H_\mathrm{tot}-\mu \hat{n}_\mathrm{tot}\right)}\right]}{\mathrm{Tr}\left[e^{-\beta \left(H_{\mathrm{tot}}-\mu \hat{n}_\mathrm{tot}\right)}\right]}.
\end{equation}
The analytic continuation result exactly matches the above equilibrium RDM up to second order in system-lead coupling. The proof follows exactly from the bosonic case \cite{Thingna2012}, since the number operator $\hat{n}_{\mathrm{tot}}$ commutes with the system-lead coupling $H_\mathrm{SB}$. In nonequilibrium, since there is no analytic form of the RDM, it is impossible to obtain a proof of validity of the MRS. Till this date, the MRS in nonequilibrium has been verified for a quantum harmonic oscillator \cite{Thingna2012,ThingnaWangHaenggi2013} and a spin-boson system \cite{Thingna2014}.

\begin{figure}[t!]
\centering
\includegraphics[width=\columnwidth]{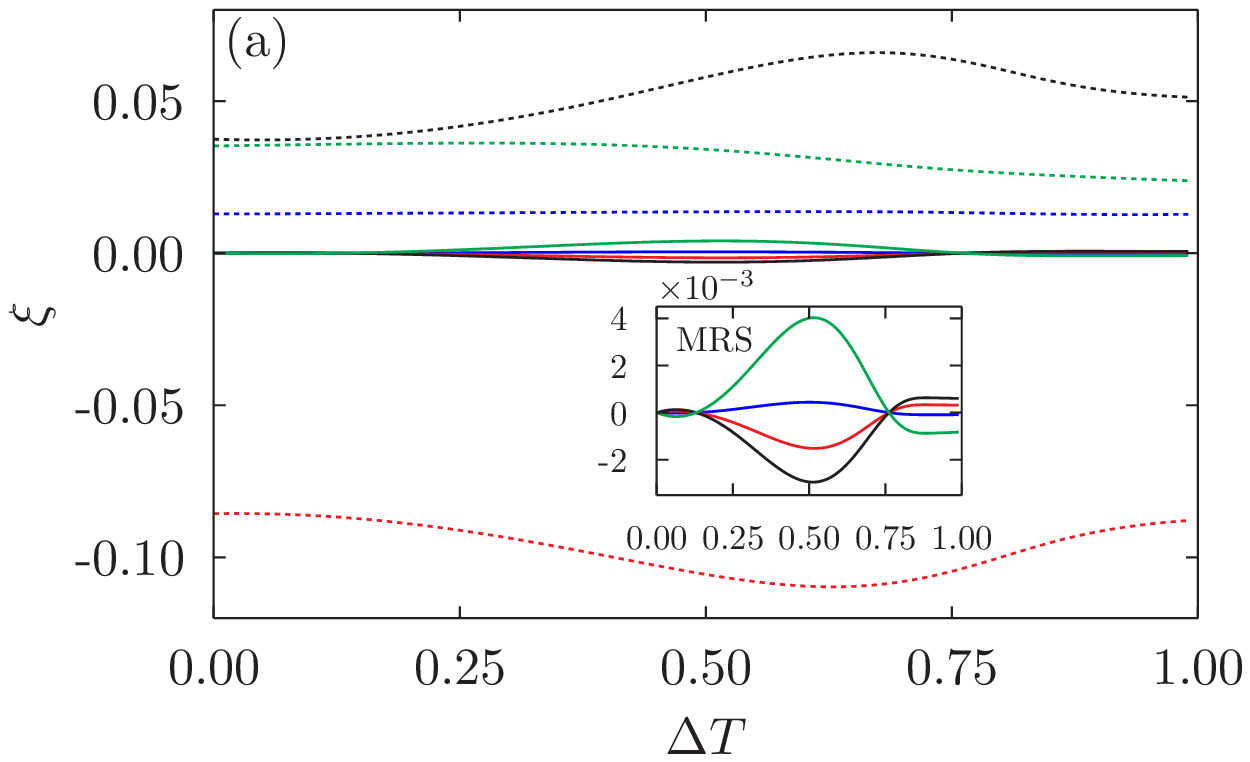}
\includegraphics[width=\columnwidth]{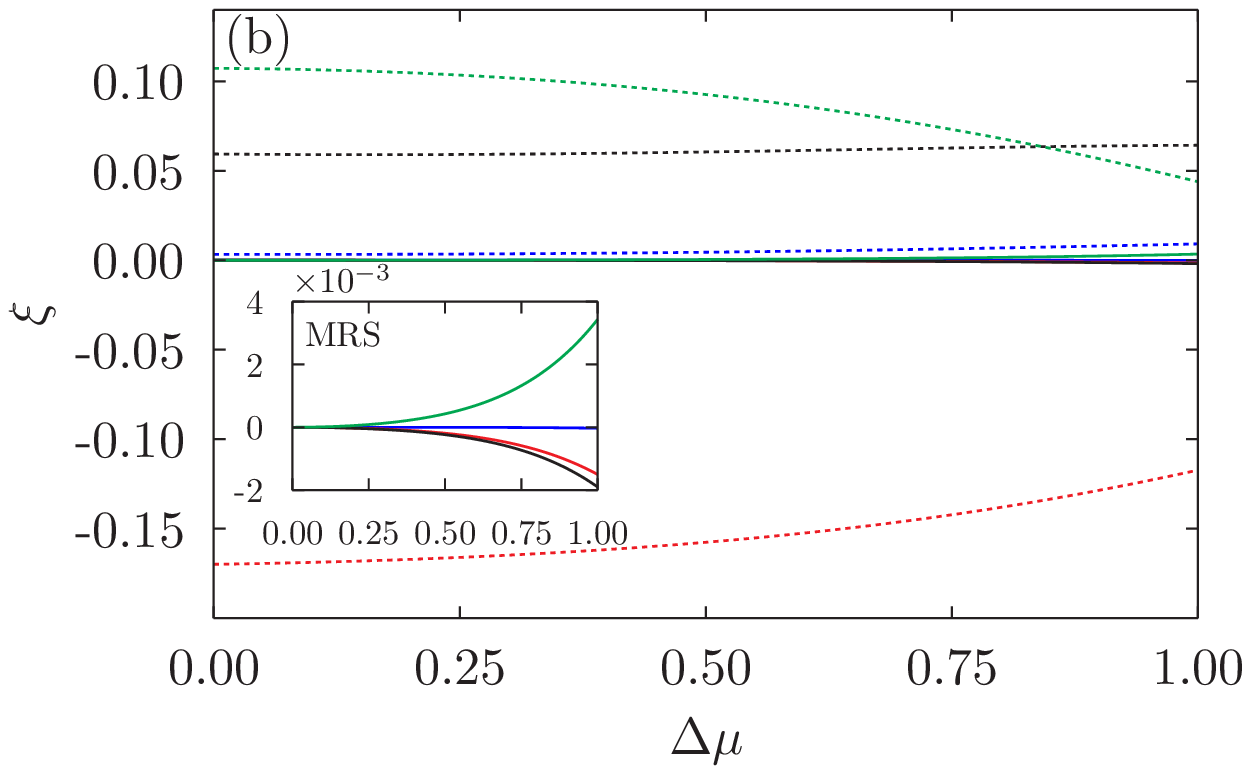}
\caption{(Color online) The exactness $\xi$ for the modified Redfield solution (solid line) and standard Redfield solution (dashed line). The off-diagonal elements $\xi_{ij}=0$ $\forall~i\neq j$ and hence we look at the diagonal elements, i.e., $\xi_{11}$ (blue), $\xi_{22}$ (red), $\xi_{33}$ (black), and $\xi_{44}$ (green). The insets show the zoom in plot only for the MRS. In (a), the chemical potential difference $\Delta\mu=0$ and the average temperature $T=0.5$, whereas in panel (b) $\Delta T=0$ and $\mu=0.5$. The left lead temperature or chemical potential is $Y_\mathrm{L}=Y(1+\Delta Y)$ [$Y\equiv \mu, ~T$] and the right lead has temperature or chemical potential $Y_\mathrm{R}=Y(1-\Delta Y)$ [$Y\equiv \mu, ~T$]. The system parameters are $\varepsilon_1=0.2$, $\varepsilon_2=0.4$, $\varepsilon_\mathrm{D}=1$, $t=1$, and $U=0$. The system-lead couplings are symmetrized such that $\Gamma_{\mathrm{L/R}}=1$.}
\label{fig:2}
\end{figure}
Next, we test the accuracy of the RDM obtained via the MRS and the Redfield master equation [Eq.~(\ref{eq:11})] for a noninteracting DQD model with the exact results available via nonequilibrium Green's function approach \cite{Dhar2012}. This can be done by defining the error $\Xi$ and the exactness parameter $\xi$ as
\begin{eqnarray}
\label{eq:20}
\Xi & = & \frac{\rho_{\mathrm{NEGF}}-\rho_{\mathrm{X}}}{\lambda^{2}},\nonumber\\
 \Xi_{ij}& = & \frac{\Delta\rho^{(0)}_{ij}}{\lambda^{2}}+\Delta\rho^{(2)}_{ij}+\lambda^{2}\Delta\rho^{(4)}_{ij}+\cdots, \nonumber \\
 \xi_{ij} &=& \lim_{\lambda^2\rightarrow 0}\Xi_{ij},
\end{eqnarray}
where X denotes the type of QME we want to test and the density matrix $\rho$ is defined in the eigenbasis. Above $\Delta\rho^{(n)}$ denotes the difference between $\rho^{(n)}_{\mathrm{NEGF}}-\rho^{(n)}_{\mathrm{X}}$ at the order $n$. The exactness parameter defined above is valid for any finite value of $\Gamma_{\mathrm{L/R}}$ and since we take the limit $\lambda^2\rightarrow 0$ (making the effective coupling $\lambda^2\Gamma_{\mathrm{L/R}}\rightarrow 0$) the results for exactness are independent of the value of $\Gamma_{\mathrm{L/R}}$. If $\rho_{\mathrm{X}}$ is accurate up to second order, $\Delta\rho^{(0)}_{ij}=\Delta\rho^{(2)}_{ij}=0$ and $\Xi \propto \lambda^{2}$. Thus, the quantity $\xi$ that is obtained from $\Xi$ in the limit $\lambda^2 \rightarrow 0$ would be an appropriate measure to test the second-order accuracy of the quantum master equation.

Figure~\ref{fig:2} depicts $\xi$ as a function of the temperature [panel (a)] and chemical potential difference [panel (b)]. When $\Delta\mu$ and $\Delta T$ equals zero we find that the function $\xi$ for the MRS (solid lines) equals zero, i.e., the MRS is exact up to second order in system-lead coupling in equilibrium. For any finite values of the affinities the dependence of $\xi$ is nonmonotonic, but in the linear response regime (near equilibrium) $\xi$ approaches zero. This clearly indicates that the MRS is strictly valid only in the linear response regime. On the other hand, the Redfield master equation (dashed lines) provides inaccurate results everywhere including the equilibrium. It is important to note here that the deviations observed are due solely to the inaccurate second-order diagonal elements and not $\rho^{(0)}$, since that would lead to a diverging $\xi$. Thus, despite its inexactness the MRS provides a crucial improvement over the standard Redfield master equation to obtain accurate solutions. The secular Redfield equation (Lindblad-type) generates $\xi$ that is the same as the Redfield (due to the negligible coherences for the chosen set of parameters) and hence not shown in Fig.~\ref{fig:2}. 
%
\section{Effect of Coulomb interaction}
\label{Sec:IV}

As shown in the previous section, the MRS provides an accurate description of the reduced density matrix of the system capturing finite dissipative effects. Next, we investigate the effect of nonlinear interactions by introducing a finite Coulomb interaction $U$ [see Eq.~(\ref{eq:2})]. Therefore, in this section, we will investigate the interplay between the Coulomb interaction, interdot tunneling strength, and finite system-lead coupling.

\begin{figure}[t!]
\centering
\includegraphics[width=\columnwidth]{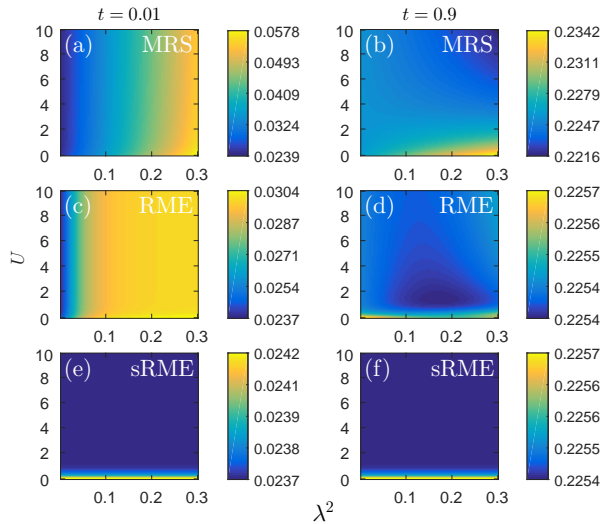}
\caption{(Color online) Color map of local population $N_1$ with varying Coulomb interaction $U$ and $\lambda^2$ for the modified Redfield solution, the Redfield solution and secular Redfield solution. The population $N_2$ (not shown) shows trends similar to $N_1$. Temperature difference $\Delta T$ is set as $0.01$ and the chemical potential difference $\Delta \mu$ is set as $0.1$. The dot energies $\varepsilon_1=\varepsilon_2=1$. The cutoff energy $\varepsilon_\mathrm{D}=1$ and the system-lead couplings $\Gamma_{\mathrm{L/R}}=1$.}
\label{fig:4}
\end{figure}

Figure~\ref{fig:4} shows the color map of the local populations [Eq.~(\ref{eq:10})] for weak (left column) and strong (right column) interdot tunneling strengths. In the weak interdot tunneling regime (Kondo impurity), the steady-state local population on each dot is low, due to the negligible exchange. In this regime, the MRS shows an increase in the local dot population as a function of the system-lead coupling strength. The RME captures this trend but the increase in the local population is smaller and faster as compared to the MRS, whereas the sRME (see Appendix B), due to the omission of the fast rotating terms, completely fails to capture this behavior and remains constant with system-lead coupling. 

Since the interdot tunneling strength between the dots is weak the dominant effect comes from the system-lead coupling. The lead can enhance or diminish the single and double occupation probabilities of the system. The rates at which the leads enhance these probabilities depend on the transition rates Eq.~(\ref{eq:13}). The enhancement rates are always larger than the diminishing rates. Thus, as the system-lead coupling increases the local populations ($N_1$ as shown in Fig.~\ref{fig:4} and $N_2$ not shown) of both the dots increase. The same behavior is observed also within the exact NEGF approach (zero Coulomb interaction case $U=0$). On the other hand, the local population is almost invariant with the Coulomb interaction $U$ for all the three solutions. The invariance is mainly due to the presence of a small local population that effectively makes the repulsive Coulomb interaction negligible.

When the interdot tunneling strength becomes comparable to the onsite energies the local population on each dot is higher as compared to the weak exchange regime. In this case, the Coulomb strength plays a significant role as seen from the right column of Fig.~\ref{fig:4}. The sRME resembles exactly the same trends as for the weak exchange coupling scenario and remains invariant with a change in the system-lead coupling. The local population within the RME framework increases as a function of the system-lead coupling at low $U$ and shows a non-monotonic behavior for intermediate $U$. At large repulsive strengths $U$, the RME also shows an increasing trend of the local population with the system-lead coupling strength. On the other hand, the MRS reveals a definite transition between an increasing local population at low Coulomb strengths $U$ to a strictly decreasing local population at high strengths $U$. This feature is not captured by the weak-coupling theories (sRME or RME) and shows a unique interplay between finite dissipation and repulsive Coulomb strength.

At low Coulomb interaction strengths $U$, the interdot tunneling strength dominates and enhances the single occupation probability that contributes positively to the local population (see the color scale at low $U$ in Fig.~\ref{fig:4}). In this regime, as the system-lead coupling increases the leads overall increase the local dot populations similar to the weak exchange coupling case. Thus, at low Coulomb strengths, we observe that the local population increases with the system-lead coupling strength. The repulsive Coulomb interaction forbids the double occupation of the dots. Hence at large repulsive strengths the lead can no longer enhance the double occupation probability, losing an important channel for increasing the local population. However, the reverse channel to diminish the double occupancy is still active. Thus, as the system-lead interaction increases the reverse channel becomes appreciable causing an overall decrease in the local population. Naturally, this effect cannot be captured by the weak-coupling theories since they do not fully account for the finiteness of the system-lead coupling, which is the major channel of diminishing the double occupation.

\begin{figure}[t!]
\centering
\includegraphics[width=\columnwidth]{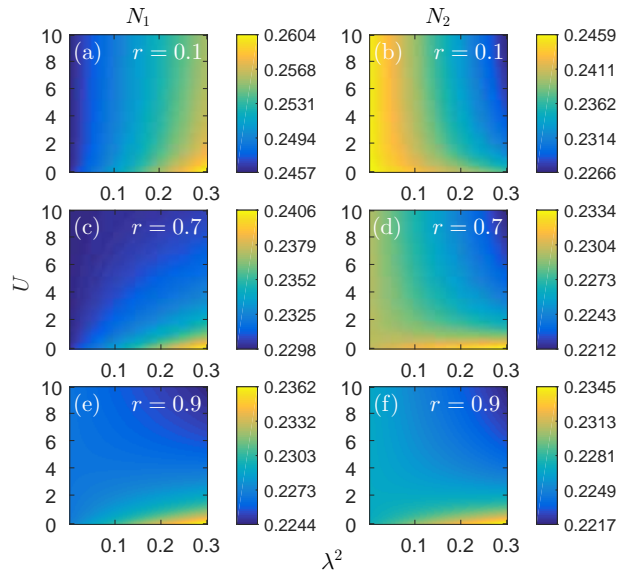}
\caption{(Color online) Color map of the local populations $N_1$ and $N_2$ with varying Coulomb interaction $U$ and $\lambda^2$ for the modified Redfield solution. The parameter $r$ induces an asymmetry with respect to the system-lead coupling with $r=\Gamma_\mathrm{R}\slash\Gamma_\mathrm{L}$. $\Gamma_{\mathrm{L}}$=1 and the interdot tunneling parameter $t=0.9$ and all other parameters are same as Fig.~\ref{fig:4}.}
\label{fig:5}
\end{figure}

To further strengthen our understanding, we consider the DQD system to be asymmetrically coupled to the two leads by setting a ratio $r=\Gamma_\mathrm{R}/\Gamma_\mathrm{L}$. In Fig.~\ref{fig:5}, we consider the case when the interdot tunneling strength $t=0.9$. If the system is weakly coupled to the right lead $r=0.1$, we obstruct the right channel that eliminates the double occupancy. Thus, the local population $N_1$ only increases as a function of the system-lead coupling for all values of the Coulomb strength. Due to the weakening of the right-lead coupling, the enhancement for the single occupation of dot $2$ is also weakened. This combined with the fact that the left lead strongly reduces the double occupancy causes the overall local population $N_2$ to reduce as a function of the system-lead coupling. It is important to note here that in the symmetric case $N_2$ resembled $N_1$ because the above processes due to the asymmetry were not present. Clearly, as we increase $r$ the system smoothly transits to the fully symmetric case, wherein the local population increase as a function of system-lead coupling for small $U$ and vice versa at large $U$.
\section{Conclusion}
In this paper, we extended the analytic continuation technique to the double quantum dot fermionic system. The accuracy of this approach was discussed by corroborating with the nonequilibrium Green's function technique. The modified Redfield solution (MRS) obtained via the analytic continuation method was numerically exact up to second order in the linear response regime. Far from linear response, the accuracy surpassed standard techniques like the Redfield or the secular-Redfield approach. Notably, the secular approximation, that is employed in several studies, failed to vary with the system-lead coupling strength wiping out all the finite coupling effects.

Due to the accurate description provided by the MRS for exactly solvable DQD system, we included the nonlinear effects of Coulomb interaction. Using the MRS we investigated the interplay between the Coulomb interaction, interdot tunneling strength, and system-lead coupling strength. The local population evaluated via the MRS revealed a complex behavior of increasing (decreasing) as a function of the system-lead coupling for weak (strong) Coulomb interactions. This switching behavior for weak and strong Coulomb interaction only occurred for strong exchange interaction between the dots and could be well explained by carefully studying the effect of the three dominant interactions on the single and double occupancy. Interestingly, the symmetric and asymmetric systems showed contrasting behaviors for the populations of the two dots elucidating the possible role of symmetry in the DQD system.

Overall, the MRS provided a stable tool in the linear response regime to study the effects of finite system-lead coupling strength in many body strongly correlated systems. This opens a whole new arena wherein one moves away from weak-coupling theories and explores the finite dissipation effects with the same ease as that of the weak-coupling approaches.
\section*{Acknowledgments}
The author thanks Hangbo Zhou for discussion. This work has been supported by FRC Grant No. R-144-000-343-112.

X. Xu and J. Thingna contributed equally to this work.

\appendix
\section*{Appendix A: Current as an observable}
In this appendix, we show that the particle current flowing through the DQD system is a local observable that depends on the system only. Moreover, such local current is only dependent on the off-diagonal elements (coherence) of the RDM. This result can be generalized to multiple sequential quantum dots with the dot number greater than two.

For the DQD model, the local particle number current $I_1$ is defined by 
\begin{eqnarray}
I_{1} &=& \left\langle i[d_{1}^{\dagger}d_{1},H_{\mathrm{S}}]\right\rangle =it\left\langle d_{2}^{\dagger}d_{1}-d_{1}^{\dagger}d_{2}\right\rangle.
\end{eqnarray}
The current that flows from the left lead \cite{ThingnaPRB2012} can be defined via
\begin{eqnarray}
I_{\mathrm{L}} =\frac{dN_{\mathrm{L}}}{dt}& =&i\left\langle[\sum_{k\in \mathrm{L}}c_{k}^{\dagger}c_{k},H_{\mathrm{tot}}]\right\rangle\nonumber\\
 & =&i\left\langle\sum_{k\in \mathrm{L}}v_{k}^{*}c^{\dagger}_{k}d_1-\sum_{k\in \mathrm{L}}v_{k}d^{\dagger}_1 c_{k}\right\rangle.
\end{eqnarray}
In steady state, there is no net change of the population on dot $1$, hence
\begin{eqnarray}
\frac{dN_{1}}{dt} & =& 0\nonumber \\
\implies I_{1}+I_{\mathrm{L}}&=&0.
\end{eqnarray}
As a result $I_{1}=-I_{\mathrm{L}}$. Similarly, we can obtain the relation for dot $2$ that $I_{2}=-I_{\mathrm{R}}$. Thus, the local particle current is indeed the current that flows through the system and is given by
\begin{eqnarray}
-I_{\mathrm{L}}=I_{1} & =&it\langle d_{2}^{\dagger}d_{1}-d_{1}^{\dagger}d_{2}\rangle \nonumber \\
 & =&it\mathrm{Tr}\left[\varrho^{\mathrm{F}}\left(d_{2}^{\dagger}d_{1}-d_{1}^{\dagger}d_{2}\right)\right]\nonumber\\
 & =&2it\mathrm{Im}\left(\varrho_{23}^{\mathrm{F}}\right),
\end{eqnarray}
where $\varrho^{\mathrm{F}}$ is the RDM in the Fock basis with the basis order in the $|00\rangle$, $|10\rangle$, $|01\rangle$, and $|11\rangle$ with $0$ and $1$ representing the empty and occupied states respectively. Therefore, a globally defined particle current can be reduced to local current for a DQD model. 

The transformation matrix $\bm{\Lambda}$ from the Fock basis to the energy eigenbasis is given by
\begin{eqnarray}
\bm{\Lambda}= \begin{pmatrix}1 & 0 & 0 & 0 \nonumber\\
0 & \Lambda_{22} & \Lambda_{23} & 0\nonumber\\
0 & \Lambda_{32} & \Lambda_{33} & 0\nonumber\\
0 & 0 & 0 & 1
\end{pmatrix}.
\end{eqnarray}
Thus, the off-diagonal element of the RDM in the Fock basis $\varrho_{23}^{F}$ can be rewritten in the eigenbasis and the particle current can be expressed as,
\begin{eqnarray}
I_{1} & =2it\mathrm{Im}\left(\Lambda_{22}\rho_{23}\Lambda_{33}+\Lambda_{23}\rho_{32}\Lambda_{32}\right).
\end{eqnarray}
Above since the diagonal elements of the RDM in the eigenbasis are real they do not contribute to the current since the local operator itself is imaginary. As a consequence, if sRME is used, since the steady-state off-diagonal elements are zero, there will always be a zero local current.

\appendix
\section*{Appendix B: Secular Redfield master equation}
The Lindblad form has been extensively used to study nonequilibrium fermionic systems. Here we derive the secular Redfield master equation and show that it has a Linblad form, preserving the positivity of the populations.

In order to apply the secular approximation, we first transform the reduced density matrix in the interaction picture or a rotating frame \cite{Blum2012} as,
\begin{eqnarray}
\rho_{nm}=\rho^\mathrm{I}_{nm}e^{-i\Delta_{nm}t}.
\end{eqnarray}
The corresponding transformed Redfield master equation Eq.~(\ref{eq:11}) reads
\begin{eqnarray}
\frac{d\rho^{\mathrm{I}}_{nm}}{dt}=\sum_{i,j}\left(\mathcal{R}^{ij}_{nm}+\mathcal{L}^{ij}_{nm}\right)\rho^{\mathrm{I}}_{ij}e^{-i(\Delta_{ij}-\Delta_{nm})t}.
\end{eqnarray}
The secular approximation assumes that the terms proportional to $\exp\left[-i(\Delta_{ij}-\Delta_{nm})t\right]$ when $\Delta_{nm}\ne\Delta_{ij}$ average to 0 and hence the terms with $\Delta_{ij}=\Delta_{nm}$ survive. Transforming back to the Schr\"{o}dinger picture, the equation decouples into the diagonal and off-diagonal parts:
\begin{eqnarray}
\label{eq:sdiag}
\frac{d\rho_{nn}}{dt}&=&\sum_{i}\left(\mathcal{R}^{ii}_{nn}+\mathcal{L}^{ii}_{nn}\right)\rho_{nn},\\
\label{eq:soffdiag}
\frac{d\rho_{nm}}{dt}&=&-i\Delta_{nm} \rho_{nm}+\left(\mathcal{R}^{nm}_{nm}+\mathcal{L}^{nm}_{nm}\right)\rho_{nm},~~~(n\neq m).\nonumber\\
\end{eqnarray}
In steady state, Eq~(\ref{eq:soffdiag}) simply implies $\rho_{nm}=0$. Thus, the above form of the secular Redfield in the steady state is analytically equivalent to the equation governing the zeroth order steady-state solution Eq.~(\ref{eq:14}). Interestingly, the same form of the secular Redfield can be obtained even in the limit that the interdot coupling $t$ approaches zero.

\end{document}